\documentclass[iop]{emulateapj}

\usepackage{natbib}
\usepackage{graphicx}

\newcommand{\beq}{\begin{equation}}
\newcommand{\eeq}{\end{equation}}
\newcommand{\bea}{\begin{eqnarray}}
\newcommand{\eea}{\end{eqnarray}}

\let\oldhat\hat

\renewcommand{\hat}[1]{\oldhat{\mathbf{#1}}}

\shorttitle{A mechanism for the dependence of sunspot group tilt angles on cycle strength}
\shortauthors{E. I\c{s}\i k}

\begin{document}


\title{A mechanism for the dependence of sunspot group tilt angles on cycle strength}
\author{Emre I\c{s}\i k$^{1,2}$}
\affil{$^1$ Department of Physics, Faculty of Science and Letters, Istanbul K\"ult\"ur University, 34156, 
Bakirk\"oy, Istanbul, Turkey \\
$^2$ Feza G\"ursey Center for Physics and Mathematics, Bo\u{g}azi\c{c}i University, 34684, 
\c{C}engelk\"oy, Istanbul, Turkey}

\email{e.isik@iku.edu.tr}

\begin{abstract}
The average tilt angle of sunspot groups emerging throughout the solar cycle determines the net magnetic flux crossing the equator, which is correlated with the strength of the subsequent cycle. I suggest that a deep-seated, non-local process can account for the observed cycle-dependent changes in the average tilt angle. Motivated by helioseismic observations indicating cycle-scale variations in the sound speed near the base of the convection zone, I determined the effect of a thermally perturbed overshoot region on the stability of flux tubes 
and on the tilt angles of emerging flux loops. 
I found that 5-20 K of cooling is sufficient for emerging flux loops to reproduce the reported amplitude of cycle-averaged tilt angle variations, suggesting that it is a plausible effect 
responsible for the nonlinearity of the solar activity cycle. 
\end{abstract}


\maketitle

\section{Introduction}
\label{sec:intro}

One of the unsolved problems of the solar activity cycle is the physical nature of the 
mechanism(s) underlying the observed variations in cycle amplitude \citep{char10}.  
Among several possibilities, reduction of poloidal flux generation by reducing the 
average tilt angle of bipolar magnetic regions has recently been considered 
as a plausible candidate. 
Analysis of tilt angle data from Mt. Wilson and Kodaikanal observatories 
between solar cycles 15-21 by \citet{dasi10} has led to the discovery that the 
cycle-averaged sunspot group tilt angle was inversely correlated with the cycle strength. 
In terms of the Babcock-Leighton dynamo process, this means that the surface source for 
the poloidal field becomes weaker for stronger cycles, potentially limiting the strength of the 
next cycle. 

A possible explanation for the observed anti-correlation is based on the effective reduction of 
the tilt angle by inflows toward activity belts, which are observed by local 
helioseismic 
techniques \citep{gh08}. Incorporation of such inflows into surface flux transport models 
has shown the efficiency of this mechanism in limiting the solar axial dipole moment 
\citep{jicss10, cjss10, cs12}. 

As already discussed by \citet{dasi10}, systematic 
changes in the tilt angle can also be led by changes in the internal structure of the lower 
convection zone, a potential location for the origin of magnetic flux loops which produce 
sunspot groups. An observational hint came from global 
helioseismology of low-degree oscillation modes by \citet{bb08}, who found a statistically 
significant reduction in the acoustic wave speed near the base of the convection zone between  
the minimum and maximum of cycle 23. 
A temperature perturbation mainly in the same 
direction (cooling) was predicted by \citet{rempel03}, who considered a magnetic layer near 
the base of the convection zone and 
obtained time-dependent solutions for radial heat transport by including radiative heating 
from below, in the presence of an imposed horizontal magnetic field reaching $10^5$~G. 

In addition to radiative effects on stratification, stronger cycles possibly involve more 
frequent flux tube explosions in the midst 
of the convection zone \citep{fmi95, rempelmsch01,hotta12}. This can also lead to a 
decrease in the radial entropy gradient, hence a decrease in 
the (negative) superadiabaticity in the lower convection zone. 

In both the convection quenching and the entropy mixing scenarios, the 
convection zone would be increasingly stabilized for stronger magnetic fields. 
Consequently, the critical field strength for the onset of flux tube instability 
would be raised with the cycle strength. Flux tubes would then become unstable at 
higher field strengths, emerge at the surface with smaller tilt angles owing to stronger 
tension force. 
 
Motivated by the helioseismic observations and the theoretical arguments summarized 
above, I determine the variation in the thermal perturbation 
required to account for the observed changes in the cycle-averaged tilt angle. 
The results indicate that a thermodynamic cycle in phase with the activity cycle at the 
base of the convection zone can be responsible for the nonlinear saturation of the 
solar dynamo.

\section{The model}
\label{sec:model}

I use a one-dimensional stratification model of the solar convection zone
\citep{ss91}, which uses the non-local mixing length formalism of \citet{ss73}. This model 
allows for a weakly subadiabatic lower convection zone below $0.775 R_\sun$, 
extending down to where the convective heat flux changes sign at 
about $0.736 R_\sun$. The convective overshoot region extends from this location 
down to $0.721 R_\sun$, with a thickness of about $10^4$~km. 

\subsection{Perturbations to the stratification}
\label{ssec:pert}

To approximate the effect of radiative heating of a magnetic layer 
in the overshoot region as estimated by \citet{rempel03}, I model the 
change in the stratification simply as a decrease in the temperature with 
an asymmetric piecewise Gaussian perturbation of the form 
\beq
T_1 = T_{\rm m}\exp\left[\frac{-(r-r_p)^2}{\sigma_\pm^2}\right],
\label{eq:t1}
\eeq
where $T_{\rm m}$ is the amplitude of the perturbation, centered at 
$r_p=5\times 10^{10}$~cm ($0.718 R_\sun$), and $\sigma_\pm$ is the characteristic width of the distribution, 
with $\sigma_-=400$~km for $r<r_p$, and $\sigma_+=4000$~km for $r\geq r_p$. 
Denoting the background thermodynamic variables by index 0 and the perturbations 
by index 1, I assume that the perturbations satisfy hydrostatic equilibrium, 
\beq
\frac{dp_1}{dr} = -\rho_1 g. 
\label{eq:hydro}
\eeq
For linear perturbations the ideal gas relation takes the form 
\beq
\rho_1 = \rho_0\left(\frac{p_1}{p_0}-\frac{T_1}{T_0}\right).
\label{eq:gas}
\eeq
Using Eq.~(\ref{eq:gas}) in Eq.~(\ref{eq:hydro}), I obtain
\beq
\frac{dp_1}{dr} = -\frac{p_1}{H_{p0}}+\rho_0 g\frac{T_1}{T_0},
\label{eq:hec}
\eeq
where $H_{p0}(r):= p_0/(\rho_0 g)$ is the pressure scale height in the 
unperturbed stratification. For simplicity, I assume that 
flux tubes leading to sunspot groups have a sufficiently low filling factor within a diffuse 
background field, so that the contribution of magnetic pressure to the 
hydrostatic equilibrium is neglected against the other terms in Eq.~(\ref{eq:hec}).

The perturbation in specific entropy, $s_1$, can be determined by writing 
energy conservation in the thermodynamic notation 
\beq
s_1 \equiv ds = \frac{1}{T}\left[du+p d\left(\rho^{-1}\right)\right], 
\eeq
where $u$ is the internal energy. Writing $du$ and $d\rho$ in terms of $dT$ and 
$dp$ and expressing the differential quantities as perturbations leads to 
\beq
s_1 = c_p\left(\frac{T_1}{T_0} - \nabla_{\rm ad}\frac{p_1}{p_0}\right),
\label{eq:s1}
\eeq
where $\nabla_{\rm ad}:=(\partial\ln T_0/\partial\ln p_0)_s=1-\gamma^{-1}$ is the adiabatic 
temperature gradient. 

The most critical quantity which determines the mechanical 
stability of magnetic flux tubes in the overshoot region is the superadiabaticity 
$\delta := \nabla - \nabla_{\rm ad}$, whose perturbation reads 
\beq
\delta_1 = -\frac{H_{p0}}{c_p}\frac{ds_1}{dr}. 
\label{eq:delta}
\eeq
Using Eqs.~(\ref{eq:gas}), (\ref{eq:s1}), and (\ref{eq:delta}), the perturbation in the 
superadiabaticity is found to be 
\beq
\delta_1 = H_{p0}\left(\frac{1}{\rho_0}\frac{d\rho_1}{dr}
	 -\frac{1}{\gamma p_0}\frac{dp_1}{dr}
	 -\frac{\rho_1}{\rho_0^2}\frac{d\rho_0}{dr}
	 +\frac{p_1}{\gamma p_0^2}\frac{dp_0}{dr}\right).
\label{eq:delta1}
\eeq

\subsection{Linear stability of magnetic flux tubes}
\label{ssec:stab}

To determine the effects of the modified stratification on rising flux tubes, I first calculate
the conditions for the linear stability of flux tubes in the overshoot region, following the 
procedure described by \citet{afmsch95}, in which linear perturbations are applied on a 
toroidal flux ring in mechanical equilibrium and in spherical geometry, using the thin flux tube 
approximation. As a function of the radial location, latitude, and field strength of the flux ring, the 
fastest-growing azimuthal wave mode is obtained from 
the real parts of the complex roots of the dispersion relation, in the unstable regime, for a set of $p$, $\rho$, $g$, 
$\delta$, $H_{\rm p}$, and $\Omega$, the angular rotation speed. Differential rotation has 
been taken into account, also for Section~\ref{ssec:nonlin}, using an internal rotation 
profile $\Omega(r,\theta)$ \citep[][Eq.~23]{isik11} representing helioseismic inversions \citep{schou98}. 

\subsection{Nonlinear dynamics of magnetic flux tubes}
\label{ssec:nonlin}

To simulate the nonlinear evolution of flux tubes, I use a code developed 
by \citet{moreno86} and extended to 3D spherical geometry in the Lagrangian 
frame by \citet{cale95}. The code solves the fluid equations in ideal MHD, taking 
into account the hydrodynamic drag force and assuming isentropic evolution for the flux tube. 
The thermodynamic quantities corresponding to the radial location of each mass element 
of the tube are determined from the stratification model described above, which has 
3000 grid points over an adaptive mesh, 
spanning from $0.56R_\sun$ to the surface. The flux tube itself has periodic boundaries and 
1000 mass elements. 

Initially a flux ring is taken to be in mechanical equilibrium, which is set by neutral buoyancy 
and a prograde azimuthal flow, which balances the magnetic curvature force in the rotating frame.
Azimuthally periodic perturbations are applied, in the form of a linear combination of modes with 
azimuthal wavenumbers from $m=1$ to $m=5$, with amplitudes of the order of
$10^{-5}H_p$. For unstable, rising flux tubes, the simulations stop when the top 
portion of the tube expands to the extent that the thin flux tube approximation 
becomes inapplicable, \emph{i.e.}, when the cross-sectional radius of the tube exceeds $2H_p$. 
This occurs at a height of about $0.98 R_\sun$. 
To measure the tilt angles, the latitudinal and longitudinal distances between the preceding 
and follower legs of emerging flux loops are obtained at the same depth 
($0.97 R_\sun$). 

\section{Results}
\subsection{Effects on stratification}
\label{sec:strat}
\begin{figure}
\epsscale{1.2}
\plotone{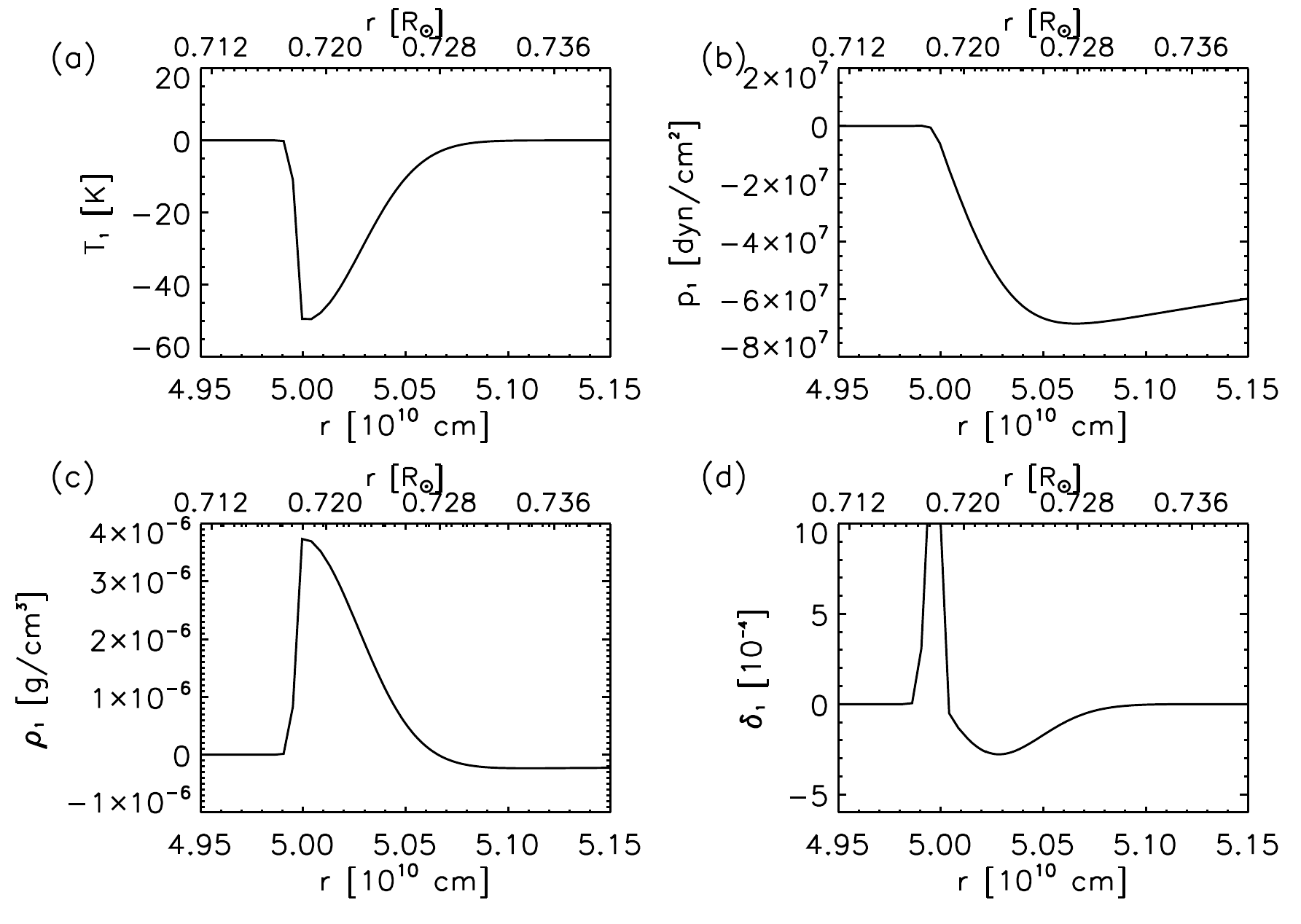}
\caption{Radial profiles of first-order perturbations in (a) temperature, (b) gas pressure, 
(c) gas density, and (d) superadiabaticity, as a function of solar radius.}
\label{fig:pert}
\end{figure}

I now solve Eq.~(\ref{eq:hec}) numerically, using a fourth-order Runge-Kutta scheme, 
by taking the equilibrium quantities as a function of 
radius from the stratification model, and the corresponding perturbations from 
Eqs.~(\ref{eq:t1}-\ref{eq:delta1}). The radial profile of the pressure 
perturbation is then obtained by setting $p_1=0$ at $r=0.56 R_\odot$ as the initial value. 
The radial profiles of the perturbations $T_1$, $p_1$, $\rho_1$, and $\delta_1$ are shown 
in Fig.~\ref{fig:pert}, for $T_m=-50$~K. Despite the simplifications made in Section~\ref{ssec:pert}, the 
resulting profile of $\delta_1$ has a similar shape and amplitude to the result 
of \citet{rempel03}. 

\begin{figure}
\epsscale{1.1}
\plotone{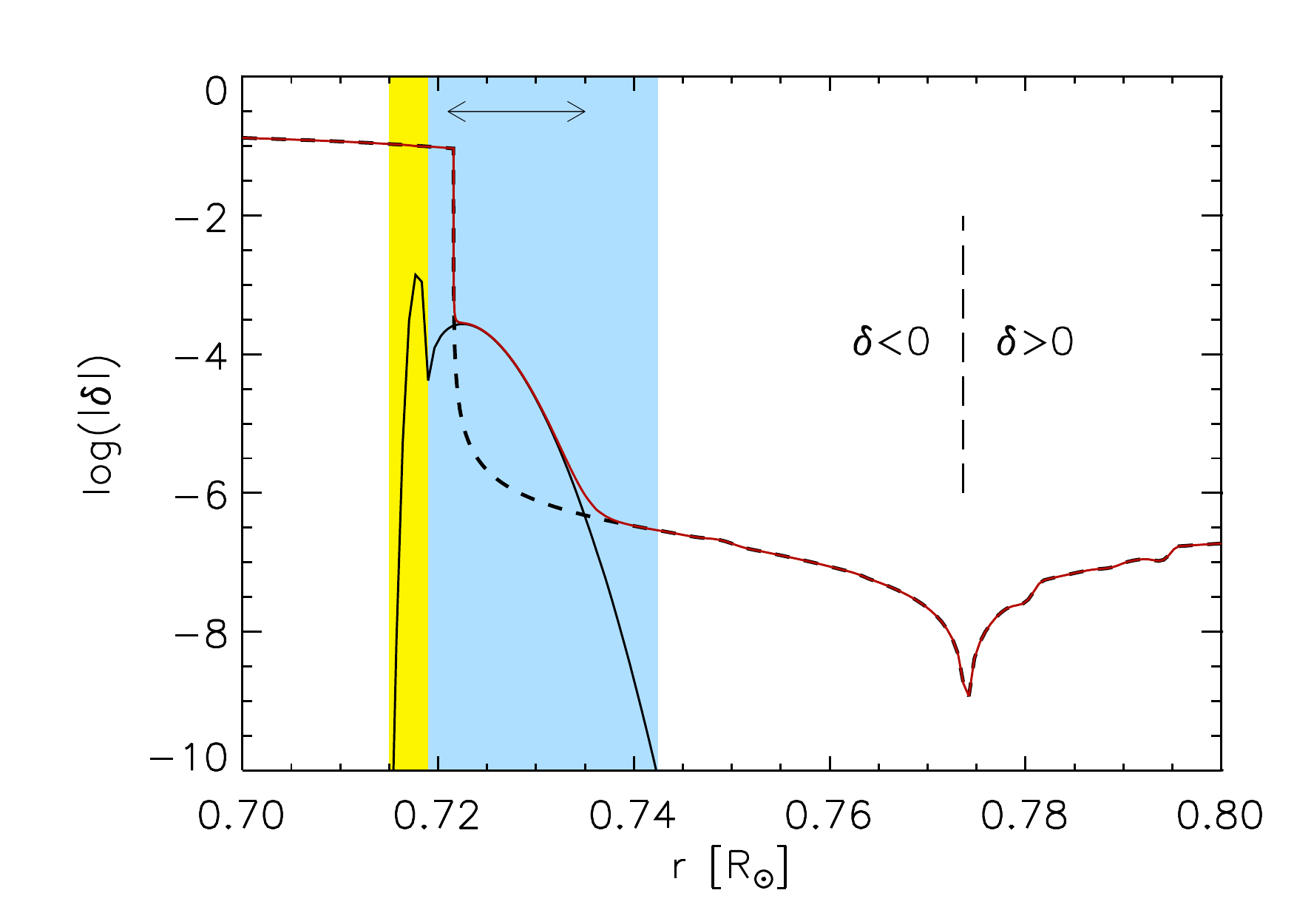}
\caption{Radial profile the absolute superadiabaticity, $|\delta|$. 
The dashed curve shows the unperturbed profile, $|\delta_0|$, with the transition 
between the subadiabatic and superadiabatic regions marked by the long-dashed 
vertical line. 
The solid black line shows the perturbation $|\delta_1|$, and the red curve shows 
$|\delta_0+\delta_1|$. The  double arrow shows the 
extent of the overshoot region. In the yellow- and blue-shaded regions 
the perturbation is positive and negative, respectively. }
\label{fig:strat}
\end{figure}

The profile $\delta_1$ of Fig.~\ref{fig:pert}d is shown in more detail in Fig.~\ref{fig:strat}, 
along with the unperturbed $|\delta|$ profile. The effect of the thermal perturbation is 
such that the stratification is destabilized within a narrow layer in the radiative zone, 
though its relative effect on the highly subadiabatic environment is insignificant 
(the yellow region). However, 
in the blue-shaded region between about $0.72R_\sun$ and $0.74R_\sun$, 
the stratification is considerably stabilized, mainly in the overshoot region (arrowed line). 

\subsection{Instability of flux tubes}

How would the modified stratification affect the mechanical stability of magnetic flux 
tubes in the convective overshoot region? I first calculate the influence of the 
thermal perturbation (Section~\ref{ssec:pert}) on the linear instability map 
of thin toroidal flux tubes subject to different strengths of thermal perturbation 
within the layer. 

\begin{figure*}
\epsscale{.45}
\plotone{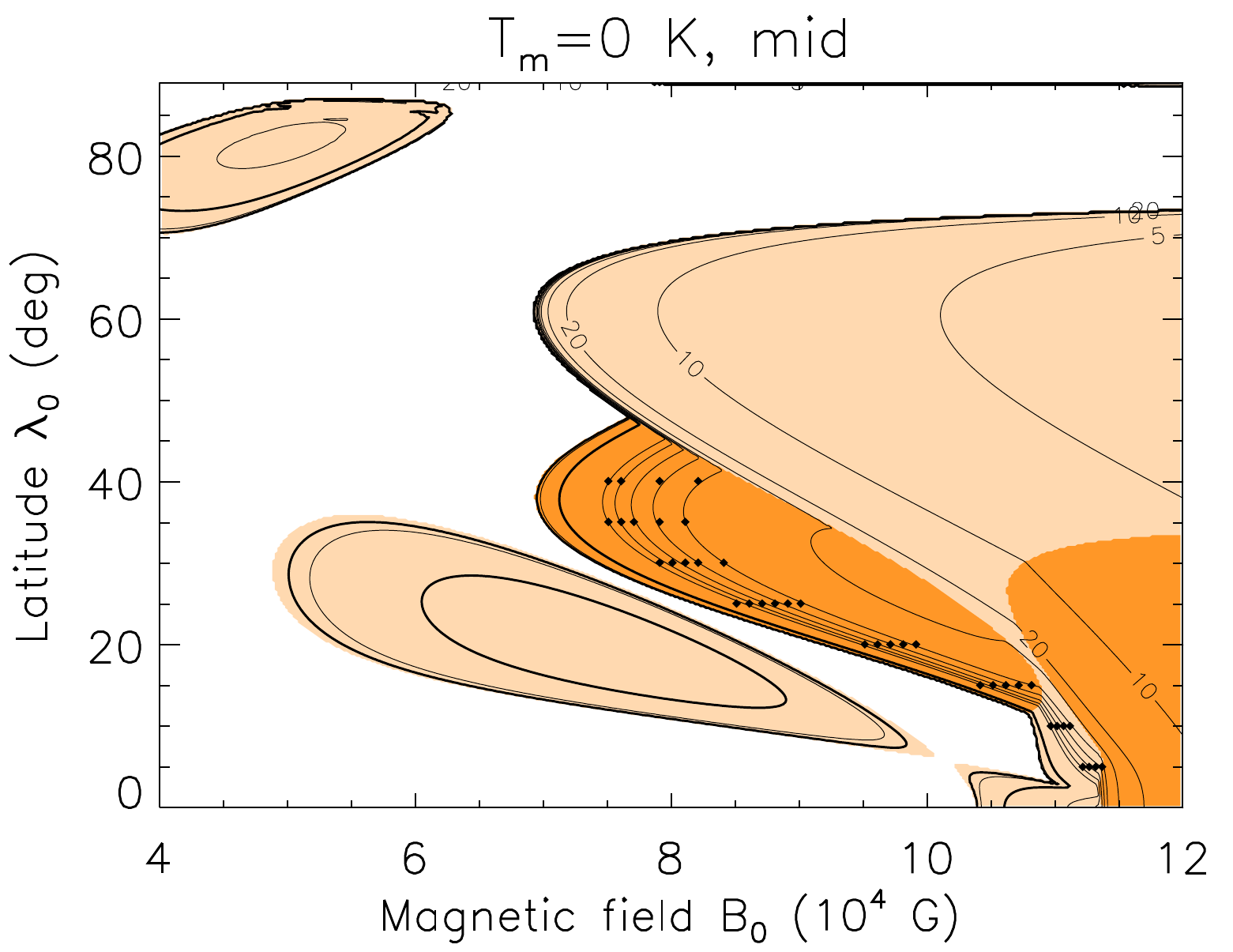}
\plotone{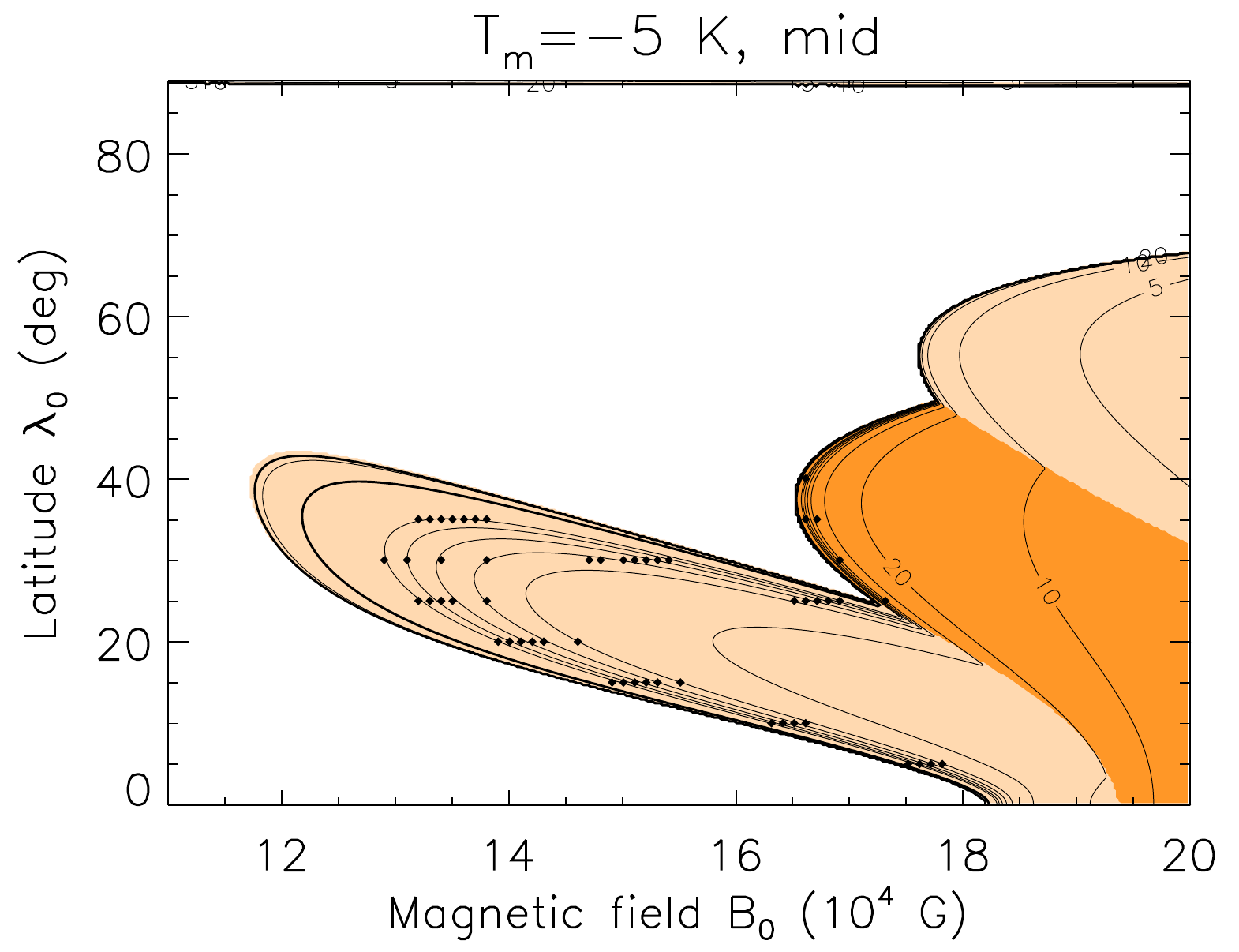} \\
\plotone{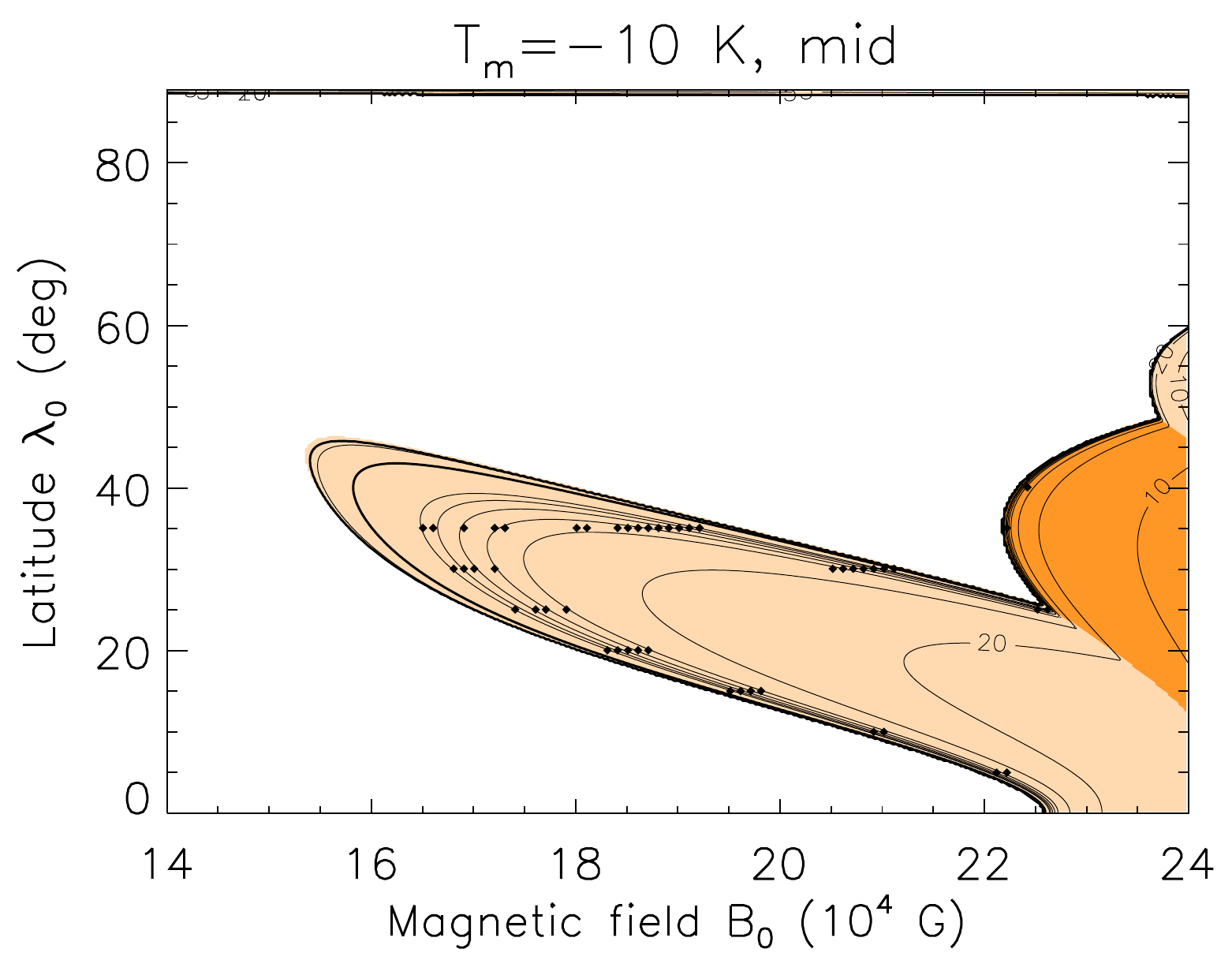}
\plotone{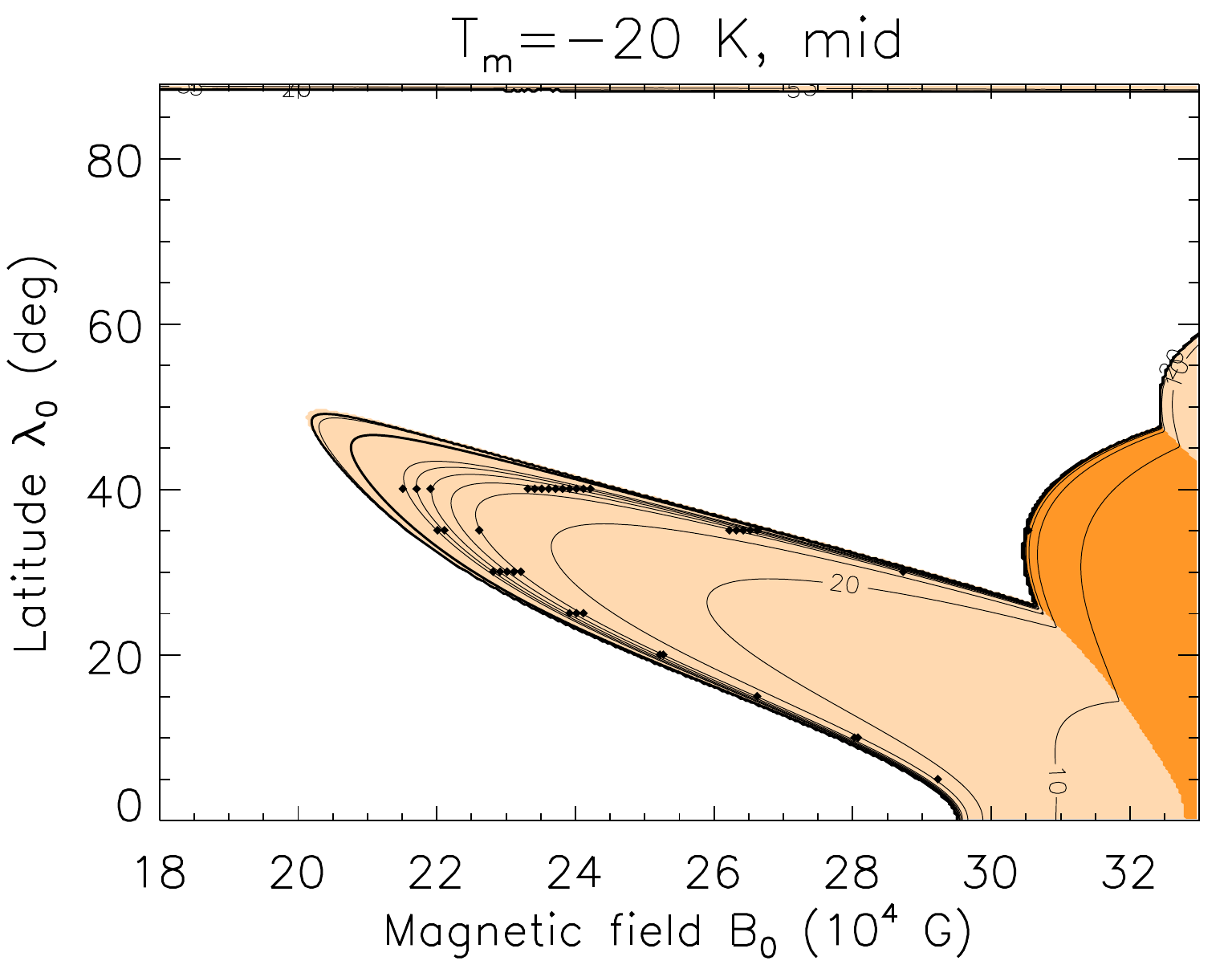}
\caption{Instability maps of a thin flux tube as a function of latitude and field strength in the 
middle of the overshoot region, for {\tt T0} and {\tt T5} (upper panels); {\tt T10} and {\tt T20} (lower panels). 
The contours show growth times from the linear stability analysis. 
The dots clustered along the densely packed 
contours (growth times 40-60 days with 5-day intervals) show the nonlinear simulations 
performed. The light (dark) shaded regions denote the wavenumber of the fastest-growing 
mode $m=1$ ($m=2$). 
It is noticeable that the instability threshold field strength shifts to larger values as the 
thermal perturbation is increased. Note that the range of field strength is different 
on each plot.
}
\label{fig:stab1}
\end{figure*}

I have set up stratification models corresponding to five values for 
the amplitude of the temperature perturbation in Eq.~(\ref{eq:t1}): $T_m=$ 0, $-5$, $-10$, 
$-20$, and $-50$~K (labeled {\tt T0}, {\tt T5}, {\tt T10}, {\tt T20}, and {\tt T50}). 
The stability diagrams resulting from the linear stability analysis (Section~\ref{ssec:stab}) 
are presented in Fig.~\ref{fig:stab1}. As $|T_m|$ is increased, 
magnetic buoyancy instability sets in at gradually higher field strengths,  
compared to the unperturbed stratification. 
For {\tt T50} (not shown here), flux tubes would 
have to be 3 to 5 times stronger to become unstable, compared to the unperturbed case, 
{\tt T0}. 

\subsection{Simulating Joy's law}
\begin{figure}
\epsscale{1.25}
\plotone{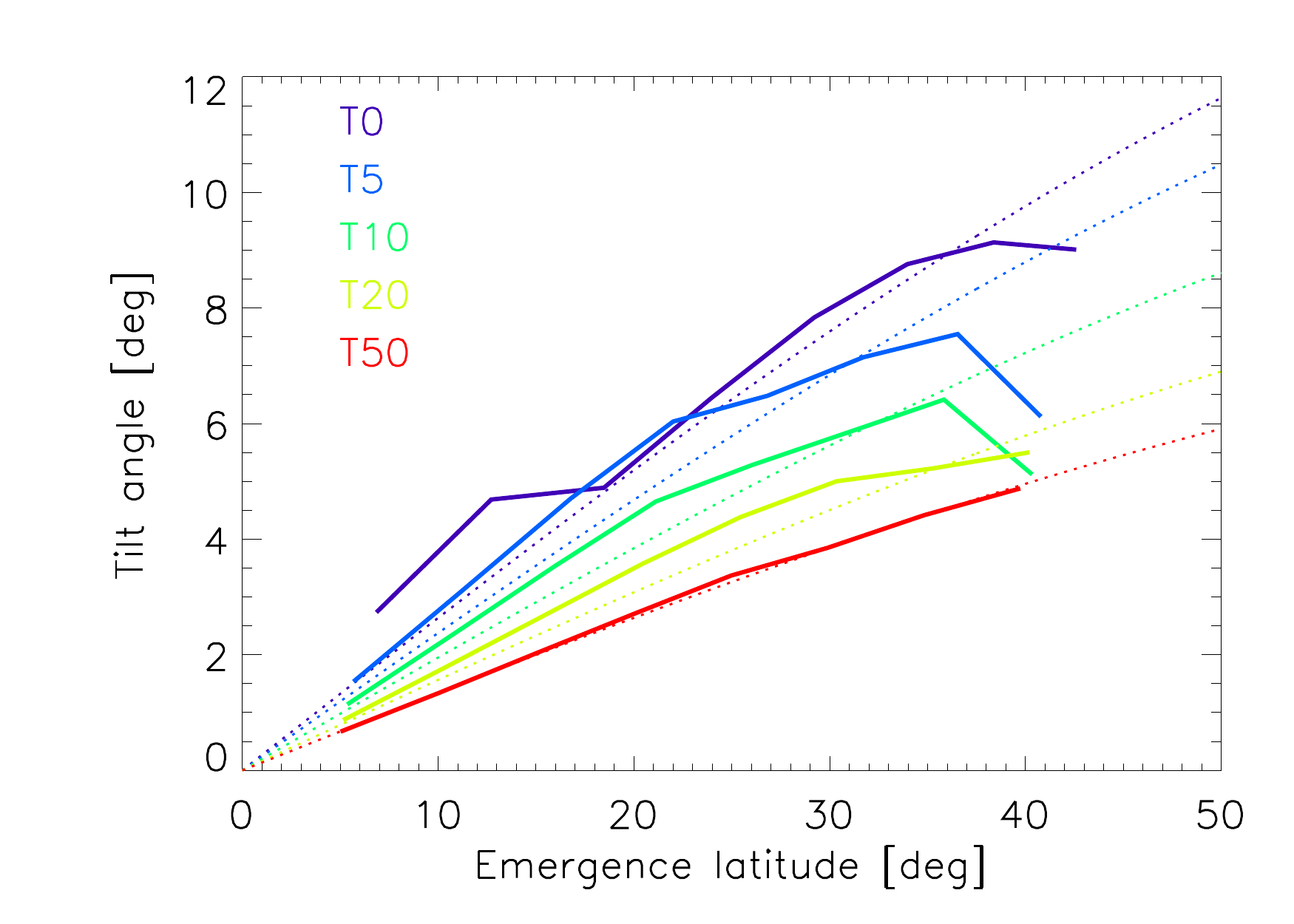}
\caption{Latitude dependence of the tilt angle (Joy's law) for simulations 
{\tt T0} to {\tt T50} with different amplitudes of local cooling. The tilt angles are averages over $5^\circ$ 
bins (continuous lines). The dotted lines show the sinusoidal fits (Eq.~\ref{eq:joy-b}). 
The average tilt angle and the steepness of the dependence decrease with increasing temperature perturbation. 
}
\label{fig:joymid}
\end{figure}

To obtain the average values and latitude dependence of the tilt angle, 
I have carried out a grid of simulations for all the cases 
{\tt T0-T50}, where the initial latitudes and field strengths of the tubes are chosen with 
$5^\circ$ intervals in latitude and for linear growth times between 40 and 60 days, 
with 5-day intervals. The initial location of the flux rings is taken at 
$0.728R_\sun$, corresponding to the middle of the overshoot region 
(same as for Fig.~\ref{fig:stab1}). 
The cross-sectional radius of the tube is set to 2000~km, which leads to a magnetic flux of 
$1.26\times 10^{22}$ Mx for a field strength of $10^5$~G. The tilt angle as a function of the 
emergence latitude (Joy's law) is plotted in Fig.~\ref{fig:joymid} for all the cases. 
To fit the simulation data, I choose the following functions, which are 
commonly used in observational studies: 
\bea
\alpha(\lambda) &=& a\lambda, \label{eq:joy-a} \\
\alpha(\lambda) &=& \gamma_0\sin\lambda,  \label{eq:joy-b}\\
\alpha(\lambda) &=& T\lambda^{1/2}  \label{eq:joy-c}, 
\eea
where $\lambda$ is the emergence latitude and $\alpha$ is the tilt angle in degrees, and 
$a$, $\gamma_0$, and $T$ are the fit coefficients corresponding to each function. 
The functions have been fitted using the nonlinear Levenberg-Marquardt algorithm. 
The form (\ref{eq:joy-a}) was used by \citet{dasi10}. 
The sinusoidal function (Eq.~\ref{eq:joy-b}) was used by 
\citet{sk12}. Their data set was based on bipolar magnetic regions from magnetograms, 
which include plage regions alongside spots, which is the possible reason for their 
systematically higher tilt angles. 
The form (\ref{eq:joy-c}) was used by 
\citet{cjss10} when fitting cycle-dependent tilt angles of \citet{dasi10}, to use in surface flux 
transport simulations. 

Joy's law coefficients resulting from the simulations are given in 
Table~\ref{tab:tilts}, which includes standard and latitude-normalized 
averages of tilt angles, the 
fitted parameters for different forms of Joy's law, and also the superadiabaticity at the initial 
location of the flux tube. The mean tilt angle and Joy's law 
coefficients are inversely proportional to the amplitude of the thermal perturbation. 
As a result of the stabilized environment, the tilt angles are systematically lower, owing to 
increasing magnetic tension between the legs of emerging flux loops. Changing the amplitude 
of cooling in the middle of the overshoot region from 5 to 20 K roughly accounts for the observed 
amplitude of cycle-averaged tilt angles for solar cycles 15 to 21. The assumption behind this 
conclusion is that the average depth from which sunspot region producing flux tubes 
originate does not change significantly as a function of cycle strength.  

\begin{table}
\begin{center}
\caption{Mean tilt angles and Joy's law parameters.\label{tab:tilts}}
\begin{tabular}{rccccrr}
\tableline\tableline 
$T_m (K)$ & $\delta$ ($\times 10^{-5}$) & $\langle\alpha\rangle$ & $\langle\alpha\rangle/\langle\lambda\rangle$ & $a$  &  $\gamma_0$  &  $T$ \\
\tableline
0	&  -0.098 & 6.69 & 0.23 & 0.25 & 15.2 & 1.39 \\
-5	&  -0.636 & 5.34 & 0.21 & 0.23 & 13.7 & 1.22 \\
-10	&  -1.16 & 4.29 & 0.17 & 0.19 &  11.2 & 1.03 \\
-20	&  -2.24 & 3.63 & 0.14 & 0.15 &  9.0 &  0.86 \\
-50	&  -54.9 & 2.91 & 0.11 & 0.13 &  7.7  & 0.72 \\
\tableline
\end{tabular}
\end{center}
\end{table}

It should be noted that taking into account the radiative heating of flux tubes has 
recently been shown to have a mild effect on Joy's law (higher slopes), in the presence of 
turbulent convective flow fields \citep{wf15}. 
In future studies, it would be of interest to include radiative diffusion in flux tube 
simulations, in conjunction with a cycle-dependent thermal perturbation. 

\section{Comparison with helioseismic evidence}

The magnitude of the change of sound speed at the base of the convection zone found in  
the helioseismic analysis of \citet{bb08} is about 
$\delta c^2/c^2 = (7.23\pm2.08)\times 10^{-5}$, expressed as the difference in the squared 
sound speed between the solar minimum 
and maximum, normalized to the minimum value. Assuming that the reduction in wave speed is 
solely due to a temperature drop, the corresponding cooling amplitude amounts to about $-150\pm 45$~K. 
Following the approach taken in \citet{bb08} and assuming that the change in the 
sound speed between cycle minimum and maximum is purely due to the change in the local 
Alfv\'en speed, an estimate for the magnetic field strength reads from 
\bea
B \simeq \left( 4\pi\gamma p \frac{\delta c^2}{c^2}\right)^{1/2},
\label{eq:bmax} 
\eea
which is about $3.6\times 10^5$~G, using the sound speed perturbation from the helioseismic 
result, and the local gas pressure from the structure model used here. 
From another perspective, \citet{rempel03} found that when a magnetic field of $10^5$~G 
quenches the convective heat conductivity by a factor of 100, a local cooling of about 40 K 
would be reached within 6 months, and 50 K within a few years, yielding a profile similar to 
Fig.~\ref{fig:pert}a. For a quenching factor of $10^4$, he found nearly 200~K within 6 months. 

The base location of the convection zone is different in the model used here ($0.736 R_\sun$) 
and the helioseismic inversions \citep[$0.713 R_\sun$,][]{ba97}. However, the instability and eruption 
properties of flux tubes in the overshoot region are most 
sensitively determined by the local superadiabaticity. Therefore, when a helioseismically 
calibrated stratification \citep[e.g.,][]{zhang12, cd11} is used, the 
superadiabaticity values given in Table~\ref{tab:tilts} would lead to similar tilt angles, 
but the initial depth of flux tubes would be shifted inwards, leading to a similar 
inverse correlation with the perturbation strength.

In the simulations presented in the previous section, $r_p$ in Eq.~(\ref{eq:t1}) 
has been taken at the radiative zone boundary ($0.718 R_\sun$), where $\delta$ changes 
very steeply (Fig.~\ref{fig:strat}). 
I have made this choice to represent the conditions set up in the model of \citet{rempel03}. 
However, the central location of the sound speed reduction observed by \citet{bb08} 
is at the base of the convection zone, which is at $0.736 R_\sun$ in the stratification 
I have used here. 
Unfortunately, the detailed radial profile or even the extent of the sound speed perturbation 
cannot be drawn from the helioseismic signal, owing to limited resolution provided by 
the low frequency waves used to probe the region. Taking $r_p$ at $0.736 R_\sun$, 
one finds that $T_m$ on the order of $-100$~K results in a thin superadiabatic layer, 
whose thickness is determined by $\sigma_{-}$ in Eq.~(\ref{eq:t1}), located in the midst of the  
stably stratified regions. This means either that the stratification where the temperature 
perturbation 
occurs would be completely restructured, or that the profile of the sound speed perturbation 
during solar maximum must have fine structure that could not be detected by 
\citet{bb08}, possibly 
involving a region of enhanced sound speed immediately below the reduction region, 
owing to excess heating from the radiative 
zone. There are indications of such a layer, though within the upper radiative zone.
Owing to such ambiguities, I have not attempted to seek a best fit to the helioseismic 
and tilt angle measurements in this preliminary study. 

\section{Conclusion}

By numerical simulations of thin flux tubes within a thermally perturbed stratification, 
I have shown that temperature variations of only 5-20 K
in the overshoot region among different cycles can reproduce the reported anti-correlation 
between the tilt angles of sunspot groups and the cycle strength \citep{dasi10}, provided that the 
emerging flux loops originate 
from about the same depth in the overshoot region in different cycles. 
Thermal fluctuations with such amplitudes among different cycle maxima are 
possible, because their magnitudes are well below the level of cooling indicated by 
the sound speed reduction from cycle minimum to maximum \citep{bb08}. 
Whether this is a relevant mechanism 
for the saturation of the solar dynamo should be tested, through $(i)$ helioseismic 
observations of sound speed variations at the base of the convection zone for 
several cycles, in conjunction with $(ii)$ measurements of the cycle-averaged tilt 
angle of sunspot groups and $(iii)$ theoretical models of the growth of the toroidal 
magnetic field and its interrelation 
with the evolving stratification \citep[e.g.,][]{cossette13}, which would be essential to explain 
the physics of the possible anti-correlation between the observed tilt angles and the 
sound speed perturbation. 

\acknowledgments
I am thankful to R. H. Cameron and M. Sch\"ussler for useful discussions which 
motivated this study. I acknowledge V. R. Holzwarth and Th. Granzer 
for help with issues related to the stratification model, M. Rempel for valuable comments, 
C. Sa\u{g}\i ro\u{g}lu for support with simulations, and the anonymous referee for 
suggestions that helped to improve the manuscript. This work has been supported 
by the Scientific and Technological Research Council of Turkey (T\"UB\.{I}TAK), under project 
grant 113F070. 


\end{document}